\documentclass{aa}
\usepackage{graphicx}
\usepackage{natbib}
\usepackage{pifont}
\pdfoutput=1
\usepackage{amsmath}

\usepackage{txfonts}
\usepackage[colorlinks=true,citecolor=blue]{hyperref}
\begin{document}

   \title{Linking radio and gamma ray emission in Ap Librae }

   %\subtitle{I. Overviewing the $\kappa$-mechanism}

   \author{O. Hervet
          %\inst{1}
          ,
          C. Boisson
          \and
          H. Sol
          }

   \institute{LUTH, Observatoire de Paris, CNRS, Université Paris Diderot, 5 Place Jules Janssen, 92190 Meudon, France\\
              \email{olivier.hervet@obspm.fr}
             }

   \date{Received November 14, 2014; accepted March 03, 2015}

% \abstract{}{}{}{}{} 
% 5 {} token are mandatory
 
  \abstract
  % context heading (optional)
  % {} leave it empty if necessary  
  {Ap Lib is one of the rare low-synchrotron-peaked blazars 
   detected so far at TeV energies. This type of source is not properly modelled by standard one-zone leptonic synchrotron self-Compton (SSC) emission scenarios.}
  % aims heading (mandatory)
   {The aim of this paper is to study the relevance of additional components that should naturally occur in an SSC scenario for a better understanding of the emission mechanisms, especially at very high energies (VHE).}
  % methods heading (mandatory)
   {We use simultaneous data from a multi-wavelength
   campaign of the Planck, Swift-UVOT, and Swift-XRT telescopes carried out
   in February 2010, as well as quasi-simultaneous data of WISE, Fermi, and H.E.S.S. taken in 2010. The multi-lambda emission of Ap Lib is modelled by a blob-in-jet SSC scenario including the contribution of the base of the VLBI-extended jet, the radiative blob-jet interaction, the accretion disk, and its associated external photon field.}
  % results heading (mandatory)
   {We show that signatures of a strong parsec-scale jet and of an accretion disk emission are present in the spectral energy
distribution.
   We can link the observational VLBI jet features from MOJAVE to parameters expected for a VHE-emitting blob accelerated near the jet base. The VHE emission appears to be dominated by the inverse-Compton effect of the blob relativistic electrons interacting with the jet synchrotron radiation. 
In this scenario, Ap Lib appears as an intermediate source between BL Lac objects and flat-spectrum radio quasars.
Ap Lib could be a bright representative of a specific class of blazars, in which the parsec-scale jet luminosity is no more negligible compared to the blob and contributes to the high-energy emission through inverse-Compton processes.}
  % conclusions heading (optional), leave it empty if necessary 


   \maketitle
%
%________________________________________________________________

\section{Introduction}

Blazars are a distinctive class of active galactic nuclei (AGN) that are known to
exhibit rapid variability, to lack bright emission lines in their
spectra, to show strong and variable polarization at optical and radio
wavelengths, and flat high-frequency radio spectra \citep{LBL_1}. They are known to emit radiation across the entire electromagnetic spectrum. 
Their broad-band continuum energy distribution (SED), from radio to 
gamma-rays, consists of two distinct 
broad bumps: a low-energy component from radio through UV or X-rays, and a 
high-energy component from X-rays to gamma-rays. In most standard
leptonic scenarios, the first bump
is interpreted as the synchrotron  emission of highly
relativistic electrons in magnetized plasmas, the second high-energy
component as due to the inverse-Compton emission from electrons
upscattering synchrotron photons (synchrotron-self-Compton scenario, SSC) or
photons from the ambient fields (external inverse-Compton, EIC).

Blazars can be classified according to the frequency 
of the peak of the low-energy (synchrotron) SED component:
\begin{itemize}
\item Low-synchrotron-peaked
(LSP) blazars, consisting of flat-spectrum radio quasars (FSRQs) and 
low-frequency-peaked BL Lac objects (LBLs), have their synchrotron peak in the infrared regime, at $\nu_s \le 10^{14}$~Hz. 
\item Intermediate-synchrotron-peaked (ISP) blazars, including LBLs and intermediate BL Lac objects (IBLs), have their 
low-energy peak at optical -- near-UV frequencies at $10^{14} \, {\rm Hz}< \nu_s \le 10^{15}$~Hz.
\item High-synchrotron-peaked (HSP) blazars,
which are almost all high-frequency-peaked BL Lac objects (HBL), have
their synchrotron peak at X-ray energies with $\nu_s > 10^{15}$~Hz.
\end{itemize}

The majority of blazars  (almost 80\%) detected at VHE belong to the HBL subclass. Their average SED as well as individual flares can in most cases be well described by simple SSC scenarios. However, a few cases of LBL and IBL sources, among which Ap Lib (PKS 1514-241), have been seen  at VHE as well and raise specific questions.

Ap Lib was first quoted as a variable star by \cite{ashbrook} before it was identified as the optical counterpart of the radio source PKS~1514-241 \citep{bond,biraud}. It had also previously been
identified with a faint, compact galaxy \citep{parkes}. \cite{LBL_1} and \cite{LBL_2} first suggested that Ap Lib was a member of the BL Lac class of objects. To date, the best available measurement that demonstrates its very fast optical variability gives a variation rate of up to $0.06 \pm 0.01$ mag/hr in the V band \citep{opt_var}. Optical polarization as high as $8.0\%$ is observed \citep{Stickel_1994}, but the average degree of polarization is quite low in the IR \citep{infrared_polarization}.

Faint emission and absorption lines confined to the nucleus were detected, which led to a first estimate of a redshift of z=0.049 \citep{Disney_1974}, which is
confirmed by the final redshift release of the 6dF
Galaxy Survey \citep{redshift}. A
supermassive black hole mass of $10^{8.4\pm{} 0.06} M_{\odot}$ was
deduced from velocity dispersion measurements together with an
elliptic host galaxy mass of $10^{11.40\pm{} 0.03} M_{\odot}$ by
\cite{Woo}. In the same paper, Ap Lib, known as an RBL (radio-selected) source \citep{AplibRBL}, was named LBL for the first time.

Ap Lib presents a rather flat radio spectrum from 8.4 to 90 GHz with
an average spectral slope of $\eta = 0.76$ ($\nu F_{\nu} \propto \nu^{\eta}$) \citep{radio_slope}.
A 43 GHz polarization was measured by \cite{radio_polarization}, only one component located at $\sim 1$ mas to the core
showed a significant polarization level of $7\%$, the core itself had
a polarization level below $0.08\%$. In the same paper the radio jet morphology is described with a jet pointing toward P.A.$= 171^\circ$, within 1 mas to the core, then turning to P.A.$= 157^\circ$.
More recently, \cite{MOJAVE} reported direct VLBI measurements of jet
flow kinematics and magnetic field properties.
\footnote{\url{http://www.physics.purdue.edu/astro/MOJAVE/sourcepages/1514-241.shtml}}

Ap Lib was first detected in X-rays by the Einstein X-ray
Observatory \citep{EXRO}. The high-energy (HE) emission of the source
(E $>$ 100 MeV) was detected by different atmospheric balloons
\citep{frye}, but only in 1999 it was realized, thanks to EGRET on the CGRO
satellite, that the gamma-ray source is associated to Ap Lib
\citep{egret}. In 2010, the H.E.S.S. collaboration \cite{hoffman}
reported very high energy (VHE) emission (E $>$ 100 GeV) detection.

Currently, only ten other LBLs or IBLs have been detected in the TeV
energy range by IACTs, namely BL Lacertae, 3C66A, W Comae, S5
0716+714, 1ES 1440+122, PKS 1424+240, VER J0521+211 and MAGIC
J2001+435. These sources are rather peculiar, for example
BL Lacertae and VER J0521+211 could have a different categorisation
following their activity, IBL in low state and HBL in high state. The
recent observation of an extended X-ray jet by the Chandra satellite
\citep{X_jet} in addition to its VHE detection makes Ap Lib a
unique LBL specimen. Moreover, pure SSC models have difficulties to
reproduce the properties of at least three of these sources, BL
Lacertae \citep{BLLac}, 3C66A \citep{3C66A}, and W Comae
\citep{W_Comae}. Scenarios that include an additional
gamma-ray emission by an external comptonization have been favoured
for these sources.
More recently, \cite{BL_Lac_2SSC} highlighted the existence of two SSC
zones in the BL Lac SED.

Previous attempts to model Ap Lib with simple one-zone SSC were not successful \citep{Ap_lib_model,Aplib_David}. The very broad high-energy bump of Ap Lib suggests that a more complex SSC model is needed to describe the emission scenario, as we show here.\\

In {Sect. \ref{section:SED}} we describe the data we used and the host galaxy correction.
In {Sect. \ref{domaineBdelta}} we follow the method of \cite{Tavecchio} to test the relevance of simple SSC scenarios and to constrain the $B$-$\delta$ domain of magnetic field and Doppler factor.
We also show that it is difficult to fit the SED with these constraints with a pure SSC code.
In {Sect. \ref{multicomponent}} we  further develop a  multi-component SSC code based on the "blob-in-jet" scenario by \cite{mrk501} by taking into account specific spectral components due to the parsec-scale jet and accretion disk, which were negligible in the cases of HBL sources, but could play a significant role in IBL/LBL objects.
VLBI radio data are used in {Sect. \ref{radio}} to deduce additional constraints on the model that we propose and discuss in {Sect. \ref{section:modelling}}. We report   
in {Sect. \ref{section:classification}} that other sources present an SED somewhat similar to Ap Lib. They might form a specific subclass of  blazars together with Ap Lib with a significant imprint of their VLBI jet on the SED even at non-radio frequencies.  

We name the various components of the jet as follows:\\
- Extended jet: jet seen in radio and X-ray from pc to kpc scales.\\
- Pc-jet (parsec-scale jet): base of the extended jet until 100 pc.\\
- Inner jet: corresponds to the string of radio knots observed along the jet axis.

 Throughout the paper, we consider a cosmological constant H$_{0}$ = 70 km.s$^{-1}$.Mpc$^{-1}$ and $\Omega_M = 0.3$.

%__________________________________________________________________

\section{Spectral energy distribution}
\label{section:SED}

\subsection{Data}
\label{data}

Since Ap Lib is variable, simultaneous or quasi-simultaneous data
are collected to build the SED. We used the  contemporaneous data
recently published  by  \cite{giommi}, including data from Planck (from 17-02-2010 to
23-02-2010), Swift-UVOT (20-02-2010 and 22-02-2010), Swift-XRT
(20-02-2010), and  Fermi-LAT (January to March 2010).  ApLib was
observed by H.E.S.S. from 11-05-2010 to 10-07-2010
~\citep{AplibLBL}, with a spectral slope of $\Gamma = 2.45 \pm 0.20$ \citep{Aplib_David}.  No strong variability being
highlighted by Fermi observations, we consider that in spite
of the temporal gap with the multi-wavelength campaign, these data can
be used in the SED modelling.  Non-simultaneous data were used to complete the general shape of the SED, but were  not be taken into account in the modelling. We show non-simultaneous points only in the energy ranges uncovered by simultaneous data. We included archive radio data from the
NASA/IPAC Extragalactic Database (NED) \footnote{\url{http://ned.ipac.caltech.edu/}} and Swift-BAT data from the ASI Science Data Center (ASDC) \footnote{\url{http://tools.asdc.asi.it/}}. Multi-wavelength data are shown in Fig. \ref{fig:interpol_SED}. Throughout the paper, the host galaxy contribution was subtracted as described below.

\subsection{Contamination by the host galaxy}
 \label{host_gal}

Ap Lib is a nearby source, therefore its host galaxy has been resolved in the optical and near-infrared (e.g. \cite{Kotilainen_1998,Urry_2000,Pursimo,Hyvonen}). To model the SED, this contamination of the non-thermal emission has to be taken into account.

Imaging studies of BL Lacs have shown that the host galaxies have de Vaucouleurs type surface brightness profiles, as elliptical galaxies do. Moreover, R-H colours \citep{Hyvonen} agree well with evolutionary population synthesis models, which predict for low-redshift, solar metallicity ellipticals R-H$\sim$2.4 for a typical age of 13 Gyr (e.g. \cite{Fioc_1997,Bruzual_2003}). To evaluate the host galaxy contribution to the observed radiation of Ap Lib, we thus used the elliptical galaxy template spectra from PEGASE~2 \citep{pegase} to generate a synthetic spectrum for a galaxy of age 13 Gyr and mass of $10^{11.40\pm{} 0.03} M_{\odot}$ \citep{Woo}, at the distance of the source.

Then to estimate the host contribution in each UVOT and WISE filter of the SED, we used the effective radius deduced from two-dimensional R-band photometric decomposition into nucleus and host galaxy, $r_{eR} = 6.72$" by \cite{Pursimo} and the aperture correction given in Eq. (4) of \cite{Young} for a de Vaucouleurs profile of a spherical galaxy.

In all subsequent figures the host contribution has been subtracted.

\section{Constraints on a simple SSC scenario}
 \label{domaineBdelta}
 
The unusual width of the high-energy bump in the SED makes
modelling with pure SSC scenarios very challenging.  No SSC model published on this source does account for VHE data 
\citep{Tavecchio,Aplib_David}, and
other radiating components are probably needed to describe
the Ap Lib spectrum. However, before considering more complex models,
we first analyse the limits of pure SSC scenarios for AP Librae in
this section.

Under the SSC assumption, we can derive physical constraints on the source
from the SED shape and the variability properties and border the B-$ \delta $ domain for the magnetic field B and Doppler factor $\delta$.

We considered a simple stationary homogeneous one-zone SSC model
\citep{mrk501}. The SSC model describes the emitting region as a spherical plasma blob of radius R, filled with a tangled magnetic
field B, that propagates with a bulk Doppler factor $\delta$ inside the jet. The low-frequency and VHE bumps are interpreted as the
synchrotron and IC emission from a population of relativistic
leptons (e$\pm$) assigned to a single-electron population for simplicity.
The primary electron energy distribution, between Lorentz factors $\gamma_{min}$ and $\gamma_{max}$,
is parametrized by a broken power-law function, with normalization factor K
(defined as the number density of electrons at $\gamma = 1$, in units of cm$^{-3}$) and indices $n_1$ below and $n_2$ above a break energy $\gamma_{break}$. The
absorption by the extragalactic background light is taken into
account by adopting the model of \cite{EBL}. This type
of SSC model has typically eight main free parameters and thus
requires high-quality data over a wide spectral range to be well
constrained.

The size of the gamma-ray emitting zone $R$ can be derived from the
minimum variability time scale $t_{var}$ following the expression $ R = t_{var}\delta c /(1+z)$. To our
knowledge, the fastest reported variability of the source is $t_{var} = 1.3$
days. A luminosity variation of a factor
about 1.8 was observed in $\sim 24$h in the V band between 1989
March 16 and March 17, and was extrapolated to a doubling time
scale \citep{opt_var}.

\begin{figure}[h]
  \centering
  \includegraphics[width=9cm]{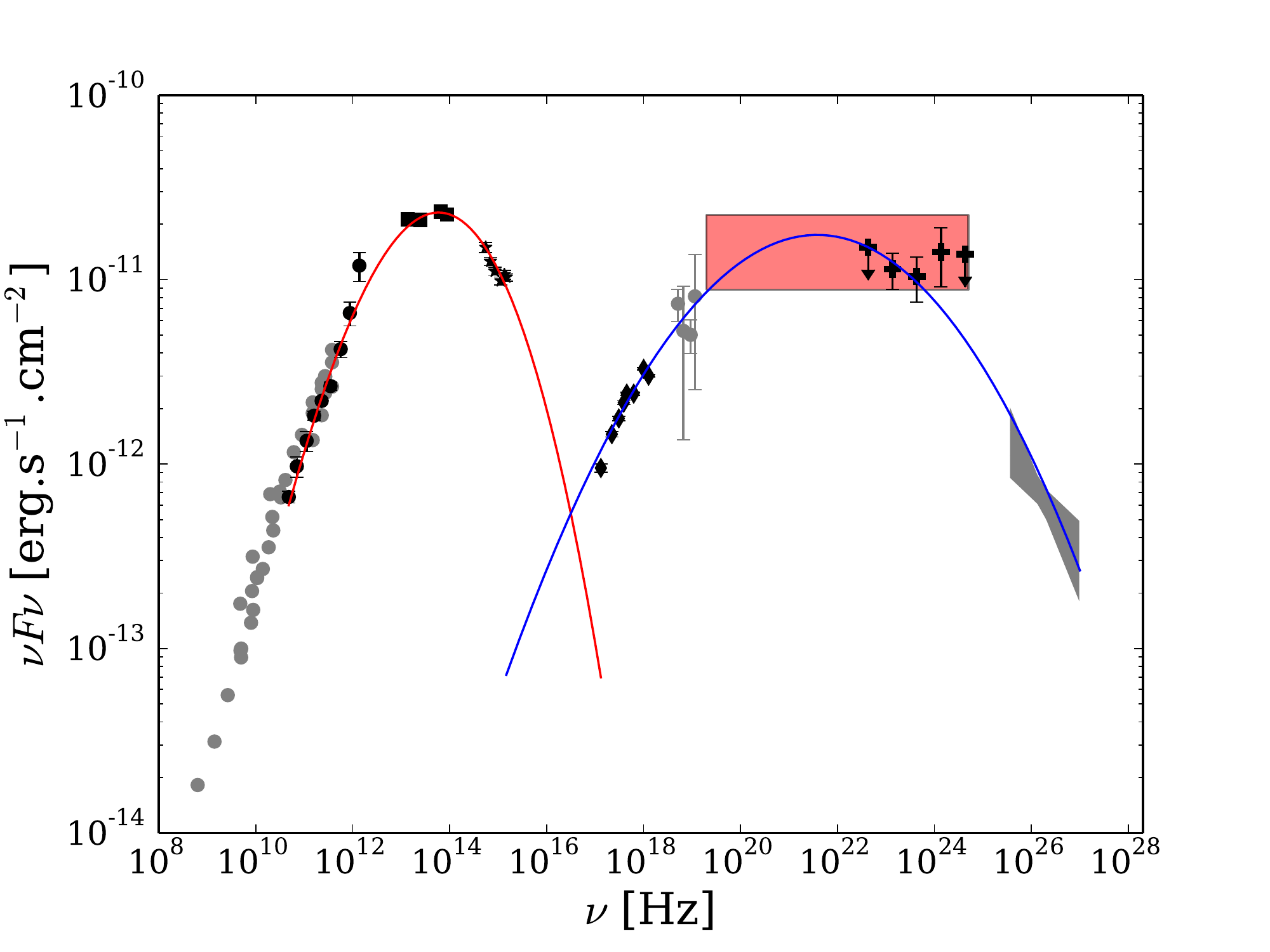}
     \caption{Interpolation of the Ap Lib SED by two-third degree 
     polynomial functions, showing 
      the synchrotron component (red) and the inverse-Compton 
     component (blue). {Black data:} Simultaneous and quasi-simultaneous 
     data, {dots:} Planck, {Squares:} WISE, {stars:} 
     Swift-UVOT, {diamonds:} Swift-XRT, {crosses:} Fermi-LAT, 
     {grey dots:} Archive data for informational
     purposes only, {grey bow-tie: } H.E.S.S..
     The {red rectangle} is the uncertainty area of 
     the Compton peak position. Only quasi-simultaneous and H.E.S.S. data are used 
     for the interpolation.}
   \label{fig:interpol_SED}
 \end{figure}

The shape of the observed SED can be used to derive observable
quantities by imposing further constraints on the physical parameters
(see e.g. \cite{Tavecchio}). The frequency and luminosity 
of the synchrotron ($\nu_s$, $L_s$) and Compton ($\nu_c$, $L_c$) peaks are estimated by interpolating the low- and high-energy bumps of the SED by a cubic polynomial function (see Fig. \ref{fig:interpol_SED}). Because the low-energy component is highly constrained by the data, the synchrotron peak is well determined. Although the
Compton component peak is poorly constrained by the available data,
it can be considered to lie between Swift-Bat and
H.E.S.S. data. Below and above the synchrotron peak, the spectrum is quite 
smooth and can be approximated with a power law  ($F_\nu \propto \nu^{\alpha}$) with
indices $\alpha_1 = 0.29$ and $\alpha_2 = 1.58,$ respectively. Adding the highest observed frequency in VHE, $\nu_{\gamma}$, Table \ref{table:observables} summarizes the eight observational constraints provided by the data.

\begin{table}[h]
\caption{\label{table:observables}Observables deduced from the SED shape.}
\centering
\begin{tabular}{lccc}
\hline\hline
\noalign{\smallskip}
Symbol & Value & Unit &\\
\hline
\noalign{\smallskip}
$L_s$                   & $1.25\times{10}^{44}$                                         & erg.s$^{-1}$.\\
$L_c$                   & $[5.1\times{10}^{43}, 1.3\times{10}^{44}]$    & erg.s$^{-1}$\\
$\nu_s$                 & $5.9\times{10}^{13}$                                          & Hz\\
$\nu_c$                 & $[2.0\times{10}^{19}, 5.0\times{10}^{24}]$ & Hz\\
$\nu_{\gamma}$  & $4.0\times 10^{26}$                                           & Hz\\
$\alpha_1$      & 0.29   &\\
$\alpha_2$      & 1.58 &\\
\hline
\end{tabular}
\end{table}

From the condition of transparency of gamma-rays to pair-production absorption, a strong constraint on the minimum Doppler factor is obtained, $\delta \geq 10.3$. For the values of the peak frequencies and slopes in Table \ref{table:observables}, the KN limit implies $\delta_{KN} < 0.29$, which is lower than the transparency limit. We can thus consider that the emission 
scenario is dominated by the Thomson regime. The synchrotron and Compton 
peak ratio gives $\gamma_b^2$. The ratio of the total luminosity of
the synchrotron peak to that of the inverse Compton is
directly related to the ratio between the radiation and magnetic
field energy density inside the emission zone. This allows
deriving a relation between the magnetic field and the Doppler factor, 
and with estimating the source dimension,
a domain of parameters such as $10.3 < \delta < 100$  and ${72.5}/{\delta^3} < B < {344}/{\delta^3}$ in Gauss.

\begin{figure}[h]
\begin{center}
\includegraphics[width=11cm]{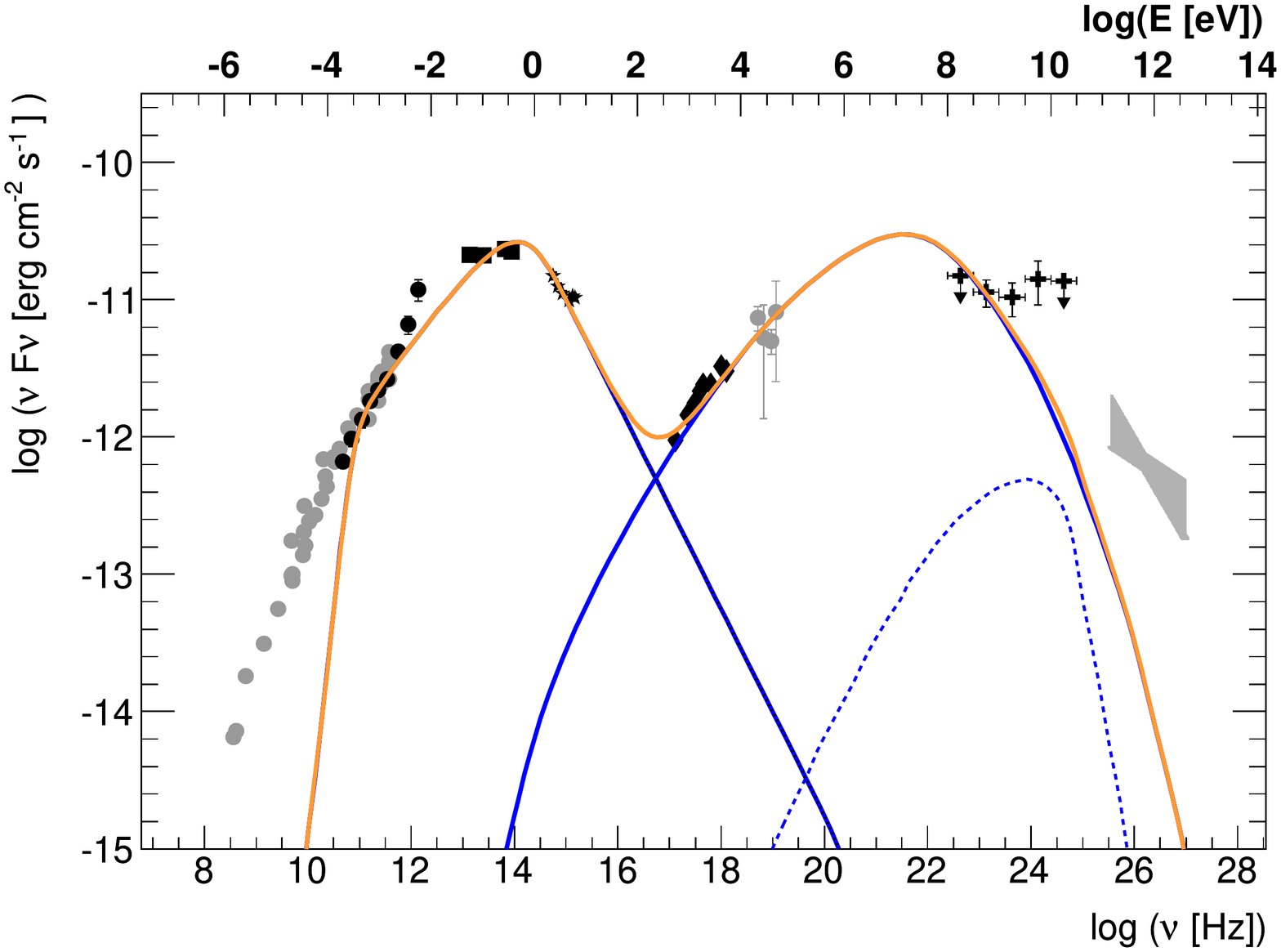}
\caption{Best tentative SSC modelling of the SED, within the B-$\delta$ constraint. {Blue lines:} synchrotron and SSC emission from the blob, {blue dashed line:} second-order SSC emission. Parameter values are given in Table \ref{table:Params_Tavecchio}. Basic SSC scenarios are unable to reproduce the VHE
fluxes (see text).}
\label{fig:SED_Aplib_tavecchio}
\end{center}
\end{figure}

Figure \ref{fig:SED_Aplib_tavecchio} illustrates the typical best fit that can be obtained by respecting 
the parameter domain B-$\delta$. The model parameter values are given in Table \ref{table:Params_Tavecchio}. The shape of the SED imposes the need for a steep 
synchrotron slope $n_2$ above the peak, so that the X-rays around $10^{17}$ Hz are not overproduced. This constraint alone prevents the Compton component from reaching the 
observed VHE flux, regardless of the choice of the other free parameters, even considering second-order scattering, where incident photons are the first-order inverse-Compton photon density.

\begin{table}[h]
\caption{Values of physical parameters used to model the SED shown in Fig. \ref{fig:SED_Aplib_tavecchio}, consistent with the constraints deduced from the observables given in Table \ref{table:observables}. {$\theta$} is the angle of the blob direction with the line of sight, {$n_1$} and {$n_2$} are the first and second slope of the electron spectrum, {$R$} is the radius of the emitting region.}
\centering
\begin{tabular}{ccc}
        \hline \hline
        \noalign{\smallskip}
    Parameter & Value & Unit\\ \hline
    \noalign{\smallskip}
    $\delta$ & $12$ & $-$\\ 
    $\theta$ & $1.0$ & deg\\ 
    $K$ & $1.5\times 10^{4}$ & cm$^{-3}$\\ 
    $n_1$ & $2.0$ & $-$\\ 
    $n_2$ & $4.5$ & $-$\\ 
    $\gamma_{min}$ & $30$ & $-$\\
    $\gamma_{max}$ & $1.0\times 10^{7}$ & $-$\\
    $\gamma_{break}$ & $5.5\times 10^{3}$ & $-$\\
    $B$ & $0.1$ & G\\
    $R$ & $1.8\times 10^{16}$ & cm\\ \hline
\end{tabular}
\label{table:Params_Tavecchio}
\end{table}

In addition, other SED features are not explained by the one-zone SSC scenario, 
such as the low-frequency radio emission, the flatness of the Fermi spectrum, and 
the change of slope in radio-mm at 250 GHz. The reduced $\chi^2$ on a simple power law fitted on radio data confirms the significance of a change of slope in radio-mm 
range ($\chi^2_{radio}$/dof = 3.7). The synchrotron bump shape does not account for the far-UV data, there is a 
clear excess above the model. This excess is usually associated with the 
thermal emission from an accretion disk.

We demonstrate in the following section by taking into account several physical 
effects that are neglected in basic SSC scenarios, but  are expected
to naturally occur in  
sources, that all of these SED features 
can be integrated in a multi-component self-consistent scenario that is able to reproduce 
the SED up to very high energies.

\section{Multi-component scenario with jet and disk}
\label{multicomponent}

After rejecting the simple one-zone SSC scenario, we focus here on the relevance of other components that can play a significant role in the emission processes.
Up to now, the influence of the extended radio jet has been mostly neglected when studying the multi-lambda emission of VHE blazars, except in a few studies where it is invoked to explain the very low frequency radio spectrum. It is commonly assumed that its contribution to the total radiative output is negligible compared to that of the VHE blob, which is historically justified by the HBLs studies in which the SSC emission from the blob dominates the SED.  However, X-ray observations of a bright extended jet by \cite{X_jet} suggest that this is not the case for the blazar Ap Lib. The two non-thermal components in the radio-mm spectrum deduced from the change of slope in the SED at 250 GHz (see Sect. \ref{domaineBdelta}) supports this view. 
One component is allocated to the blob and another to the jet. They are associated to SED peaks in the ranges [$8.82\times10^{4}$, $5.55\times10^{5}$] GHz and [$1.37\times10^{3}$, $1.36\times10^{4}$] GHz.

It is relevant to consider two populations of relativistic electrons in the non-thermal emitting zone of Ap Lib, which are related to the jet and to a blob embedded in it and contribute to the SED. In this case it becomes necessary to take into account the radiative components related to the jet particle population, mainly the jet SSC contribution and the inverse Compton of the blob particles on the jet synchrotron radiation. 
Thanks to radio VLBI data, which allow a detailed kinematic description of the jet as discussed in Sect. \ref{radio}, this new component provides some constraints on the jet parameters without additional degrees of freedom in the global picture of the source.\\

We propose a revisited blob-in-jet model, taking into account an inhomogeneous jet describing the non-thermal radiation from one peak seen in the synchrotron bump, including SSC and mutual inverse-Compton effects.
The new model implemented here is based on the work done by \cite{mrk501} and \cite{mrk421}, which was applied to HBL sources where effects related to the pc-jet were negligible. Hereafter we describe only the further developments and improvements in the scope of our study. We took into account the absorption of the VHE blob emission by pair production due to photon-photon interaction with the jet radiation, and the Comptonization of the jet radiation by blob particles. The blob moves with a Doppler factor higher than that of the jet, which induces Lorentz transformations between the blob and jet frames. 
 To complete the picture, we also include a basic description of the disk emission and scattering of its light by the broad-line region (BLR), which probably plays a role in the SED of LBL sources such as Ap Lib.

\subsection{Basic jet modelling}

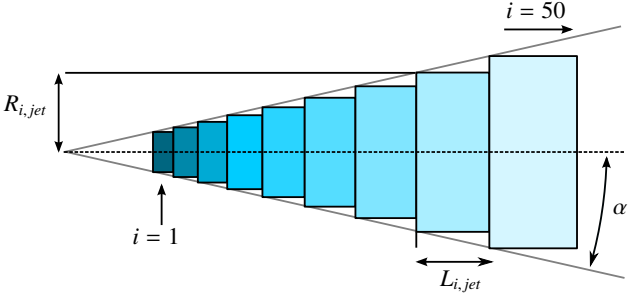
\begin{figure}[h]
\begin{center}
\includegraphics[width=8cm]{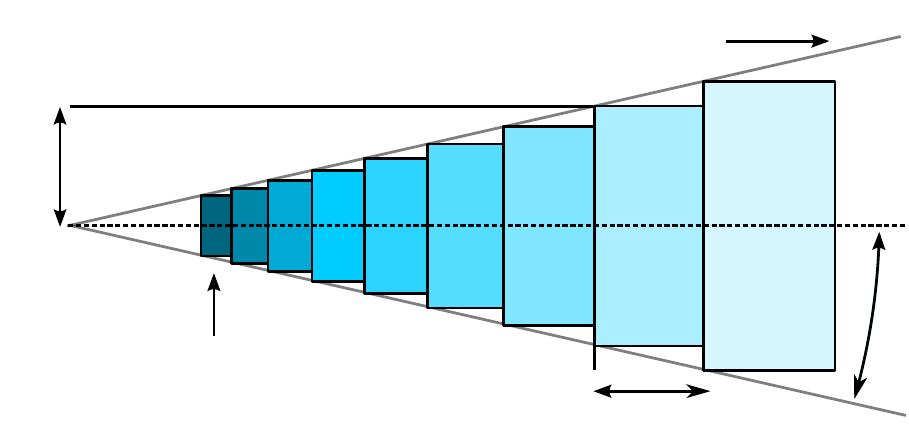}
    \put(-185,20){\color[rgb]{0,0,0}\makebox(0,0)[lb]{\smash{$i=1$}}}%
    \put(-233,70){\color[rgb]{0,0,0}\makebox(0,0)[lb]
    {\smash{$R_{i,jet}$}}}%
    \put(-45,105){\color[rgb]{0,0,0}\makebox(0,0)[lb]{\smash{$i=50$}}}%
    \put(-70,10){\color[rgb]{0,0,0}\makebox(0,0)[lt]
    {\begin{minipage}{0.10651102\unitlength}\raggedright 
    $L_{i,jet}$\end{minipage}}}%
    \put(-5,35){\color[rgb]{0,0,0}\makebox(0,0)[lt]{\begin{minipage}{0.04785274\unitlength}\raggedright $\alpha$\end{minipage}}}%
\caption{Geometry of the jet. The jet is divided into 50 slices whose radius $R_{jet}$ and width $L_{jet}$ evolve along the jet according to the opening angle $\alpha$. The {shades of blue} show the particle density and magnetic field decrease along the jet.}
\label{fig:geometry_jet}
\end{center}
\end{figure}

The jet is represented by a cone discretized in cylindrical slices (see Fig. \ref{fig:geometry_jet}). Absorption and emission coefficients are calculated for each slice. The radiation transfer is computed in the direction of the jet propagation. We express the intensity for each slice as\begin{equation}
I_{i,jet}(\nu) = \frac{j_{i,jet}(\nu)}{k_{i,jet}(\nu)}\left[1-\exp(-\tau_{i,jet}(\nu))\right]
,\end{equation}

where $\tau_{i,jet}(\nu) = L_{i,jet}k_{i,jet}(\nu)$ is the optical thickness, $j_{i,jet}(\nu)$ is the electron self-synchrotron emission coefficient, $k_{i,jet}(\nu)$  the absorption coefficient , and $\nu$  the frequency in the jet frame. Further absorption by pair production was also considered as described hereafter by Eq. \ref{gg}.

The electron density $N_{jet}$  evolves along the jet with the slice radius as
\begin{equation}
\label{eq:dens}
N_{i,jet} = N_{1,jet} \left(\frac{R_{1,jet}}{R_{i,jet}}\right)^2
.\end{equation}

According to our conical description of the jet, we can describe the expansion speed,
\begin{equation}
\frac{d(R_{i,jet})}{dt} = \beta_{jet} \tan{\alpha} = \beta_{jet} \frac{R_{i,jet}}{D_{i,jet}}
,\end{equation}

with $\beta_{jet}$ the jet velocity assumed constant, and $D_{i,jet}$ the distance of slice i to the central engine.
We consider that the pc-jet expansion is free with an expansion velocity $d(R_{i,jet})/dt$ of the order of the Alfven speed.

Then we can express the magnetic field of the slice i:
\begin{equation}
B_{i,jet} = \sqrt{4\pi N_{i,jet} m_e} \left(\beta_{jet} \frac{R_{i,jet}}{D_{i,jet}}\right)
.\end{equation}

Following Eq. \ref{eq:dens}, we can write
\begin{equation}
B_{i,jet} = B_{1,jet} \frac{D_{1,jet}}{D_{i,jet}}
.\end{equation}

This corresponds to the variation of the magnetic field along the jet as actually observed for a sample of VLBI radio jets  by \cite{B_radio_pc}.

\subsection{Blob radiation absorption by the jet}
\label{blob-jet}

We assume that $R_{blob} \ll R_{jet}$, such that the jet radiation absorption by the blob is considered negligible. However, the blob radiation absorption by the jet must be taken into account. In the high-energy range, the blob and jet photons can interact and create $e^+e^-$ pairs. For each slice we used the approximation of \cite{abs_gamma} in cylindrical geometry to estimate the pair absorption. We define the optical thickness of a jet slice $\tau_{i,\gamma\gamma}(\epsilon_c)$ by
\begin{equation}
\tau_{i,\gamma\gamma}(\epsilon_c) = 0.2\sigma_{T}\frac{1}{\epsilon_c}n_i(1/\epsilon_c) \frac{3}{4} L_{i,jet}
\label{gg}
.\end{equation}

In the source frame $V_{blob} > V_{jet}$, and according to the speed transformation law, we can express the blob velocity in the jet frame $V_{b}'$ by
\begin{equation}
V_{b}' = \frac{V_{b} - V_{jet}}{1-\dfrac{V_{b}V_{jet}}{c^{2}}}
,\end{equation}

and thus express $\Gamma_{b}'$ and $\delta_{b}'$ as a function of $V_{b}'$.\\

In the jet frame, slices in front of the blob receive a blueshifted radiation from it. We can write the blob frequencies and intensities transformations along the line of sight as $\nu_{b}' =\nu_{b} \delta_{b}'$ and $I_{b}' = I_{b} \delta_{b}'^3$ and handle the absorption of this radiation slice by slice along the jet with  frequencies and intensities of the blob expressed in the observer frame as 
\begin{equation}
\nu_{obs} = \nu'_b \frac{\delta_{b}}{\delta'_b}(1+z)^{-1}
\end{equation}
\begin{equation}
I_{obs} = I'_b \left(\frac{\delta_{b}}{\delta'_b}\right)^3(1+z)^{-1}
.\end{equation}

\subsection{External jet radiation on the blob}

The high-energy particle density in the blob is much higher than that in the jet since the so-called blob is by definition a compact zone of efficient particle acceleration processes due to shocks, turbulence, or magnetic reconnection, with a large population of relativistic electrons.
The inverse-Compton emission of the blob particles on the local jet radiation can be significant, while the inverse-Compton emission of the jet particles on the blob radiation remains negligible. To describe this effect, we calculated two radiation fields from the jet, one behind the blob and one in front of it, calculated via a radiation transfer in the direction of (resp. opposite to) the jet propagation. The sum of these two fields acts like an almost isotropic field in the blob region, and the corresponding jet frequencies and intensities in the blob frame can be simplified as follows:
\begin{equation}
\nu'_{jet} = \Gamma'_{jet} \nu_{jet}
\label{nu}
\end{equation}
\begin{equation}
I'_{jet} = \frac{1}{4\pi}\int d\Omega\delta_{jet}'^3 I_{jet} = \Gamma'_{jet} I_{jet}
\label{I}
.\end{equation}

This component of external inverse-Compton on the jet radiation due to a blob-jet interaction is found to play an important role for the VHE fluxes of Ap Lib (see Fig. \ref{fig:SED_total}).

\subsection{Disk and BLR radiation}
\label{S:disk}

As we have seen in Sect. \ref{domaineBdelta}, the UVOT spectrum suggests an additional radiating UV component. In AGNs, such a component can be most naturally attributed to the thermal emission of the accretion disk. The shape of the SED shows that the UV flux density is at least higher by a factor 10 than that of X-rays, therefore we estimate that the disk luminosity is widely dominated by the thermal bump that we approximated here by a black-body at a given temperature. 
\\

At higher energies the Fermi spectrum is rather flat. This feature cannot only be described by the blob and jet inverse-Compton components that are due to the strong constraint on the steep UV synchrotron slope (see Sect. \ref{domaineBdelta}). It suggests that an additional component radiates in the Fermi range that we can naturally identify as an additional external inverse-Compton emission.

As a result of the high blob Lorentz factor, the disk radiation interaction with particles in the blob frame is too highly redshifted to be significant, therefore we expect that this inverse-Compton emission instead originates from the BLR disk that reprocesses radiation on the blob.
To model this radiation, we approximated the accretion disk to a  point-like source ($D_{blob} \gg R_{disk}$). We assumed that the blob is located inside the BLR and propagates into an external isotropic photon field that is initially emitted by the disk, but is reprocessed by the BLR. The intensity of this field can be expressed as
\begin{equation}
I_{BLR}(\nu, T_{disk}) = \tau_{BLR} \frac{L_{disk}}{4\pi R_{BLR}^2}\frac{I_{planck}(\nu, T_{disk})}{(\sigma/\pi)T_{disk}^4}
,\end{equation}

where $\sigma$ is the Stephan-Boltzmann constant, $I_{planck}$ is the Planck intensity, and 
$\tau_{BLR}$ is a function of the reprocessing efficiency and the covering factor. Lorentz transformation of this radiation field applies in the same way as in Eqs. (\ref{nu}) and (\ref{I}).

\section{Constraints on blob and jet from VLBI radio data}
\label{radio}

We have shown the influence of the blob and  jet on the radiative output of Ap Lib. Now it is essential to constrain their physical parameters as much as possible so as not to increase the number of free parameters in the global MWL scenario. We rely for this on \cite{MOJAVE} and focus especially on the data obtained by the MOJAVE programme at 15.4 GHz on December 26, 2009, the closest epoch to the MWL observation campaign of February 2010.

\subsection{Radio core features}
\label{radio_core}

The Ap Lib radio core on December 26, 2009 shows an ellipse shape with the major axis oriented at 341$^\circ$ on the sky plane as described in the Table 4 from \cite{MOJAVE} .
The core flux density detected is $\nu F\nu_{core} = 1.52\pm 0.08\times 10^{-13}$ erg.s$^{-1}$.cm$^{-2}$ (see Fig. \ref{fig:SED_total}).

This core emission is strongly associated with the standard extended emission zone in the radio range of the SED. In our model we associated this radio core emission with the base of the simulated pc-jet that strongly emits in radio.
Assuming a conical jet description and small angle approximation, we can thus constrain the modelling parameters of the pc-jet via the radio core features. 
We assume that the half-minor axis of the ellipse corresponds to the maximal radius of the pc-jet base, which has the core flux density, namely $R_{core} = 1.4 \pm 0.6\times 10 ^{17}$cm ($D_L = 218$ Mpc). This means that we have to resolve the following equation system:
\begin{equation}
\label{eq:core}
R_{n,jet} = R_{core}\\
\&\\
\sum_{i=1}^n(\nu F_\nu)_{i,jet} = (\nu F_\nu)_{core}
,\end{equation}
where $\sum_{i=1}^n$ is the sum over the n pc-jet slices contributing to the core emission (see Fig. \ref{fig:geometry_jet}).

The solution of this equation system is strongly model dependent. To find the appropriate n slice, we adjusted the pc-jet parameters taking into account the modelling constraints given by the SED shape.
With these constraints the pc-jet space parameter is strongly reduced. When we fulfil these constraints, we can estimate the half-opening angle of the pc-jet $\alpha = 0.4^\circ$ and the size of the core emitting region $\sum_{i=1}^nL_{i,jet} = 9$ pc for a radius $R_{n,jet} = 1.9 \times10 ^{17}$cm consistent with the measured value.\\

Moreover, we assume that the major axis of the radio core ellipse $Ma_{core} = 8.6 \pm 3.4\times10 ^{17}$ cm corresponds to the size of this emission zone projected on the sky plane. With a conical jet seen at small angle, the length of the core projected is $L_{core} = Ma_{core} - R_{core} = 6.7 \pm 3.4\times10 ^{17}$cm, and so the viewing angle $\theta$:
\begin{equation}
\label{eq:theta}
\theta = \arctan\left(\frac{L_{core}}{\sum_{i=1}^nL_{i,jet}}\right) = 1.4^\circ\pm0.7^\circ
.\end{equation}

Because of the adiabatic expansion of the pc-jet, the 9 pc core
widely dominates the global emission of the total simulated pc-jet. The pc-jet emission between 9 and 100 pc contributes significantly to the global SED only in the radio range. We decided to fix the pc-jet length at 100 pc because the emission of the larger scale jet is negligible in the SED.

\subsection{Radio knot velocities}

During the observation of December 26, 2009, seven radio knots have been identified in addition to the core emission. In the Table 5 of \cite{MOJAVE}, three of these knots show a clear superluminal motion with mean apparent velocities $\beta_{app}$ of $6.03(\pm 0.34)$c, $6.41(\pm 0.19)$c, and $6.04(\pm 0.43)$c. In the following we use the average value $\beta_{app} = 6.16$ c.
 We can link the apparent velocity to the real velocity $\beta$ with
\begin{equation}
\beta = \left(\frac{\sin \theta}{\beta_{app}} + \cos \theta \right)^{-1}
,\end{equation}
where $\theta$ is the angle between the jet direction and the line of sight.\\

The Doppler factor $\delta$,
\begin{equation}
\delta = \frac{\sqrt{1-\beta^2}}{1-\beta \cos \theta}
,\end{equation}

has a mean value of $\delta = 21.8\pm ^{9.8}_{4.4}$, which is consistent with typical VHE blazar Doppler factors. Within this scheme, the radio knot velocity is similar to that of the simulated blob in the blob-in-jet VHE scenario.

\subsection{Position and size of radio knots}

\begin{figure}[h]
\begin{center}
\includegraphics[width=8cm]{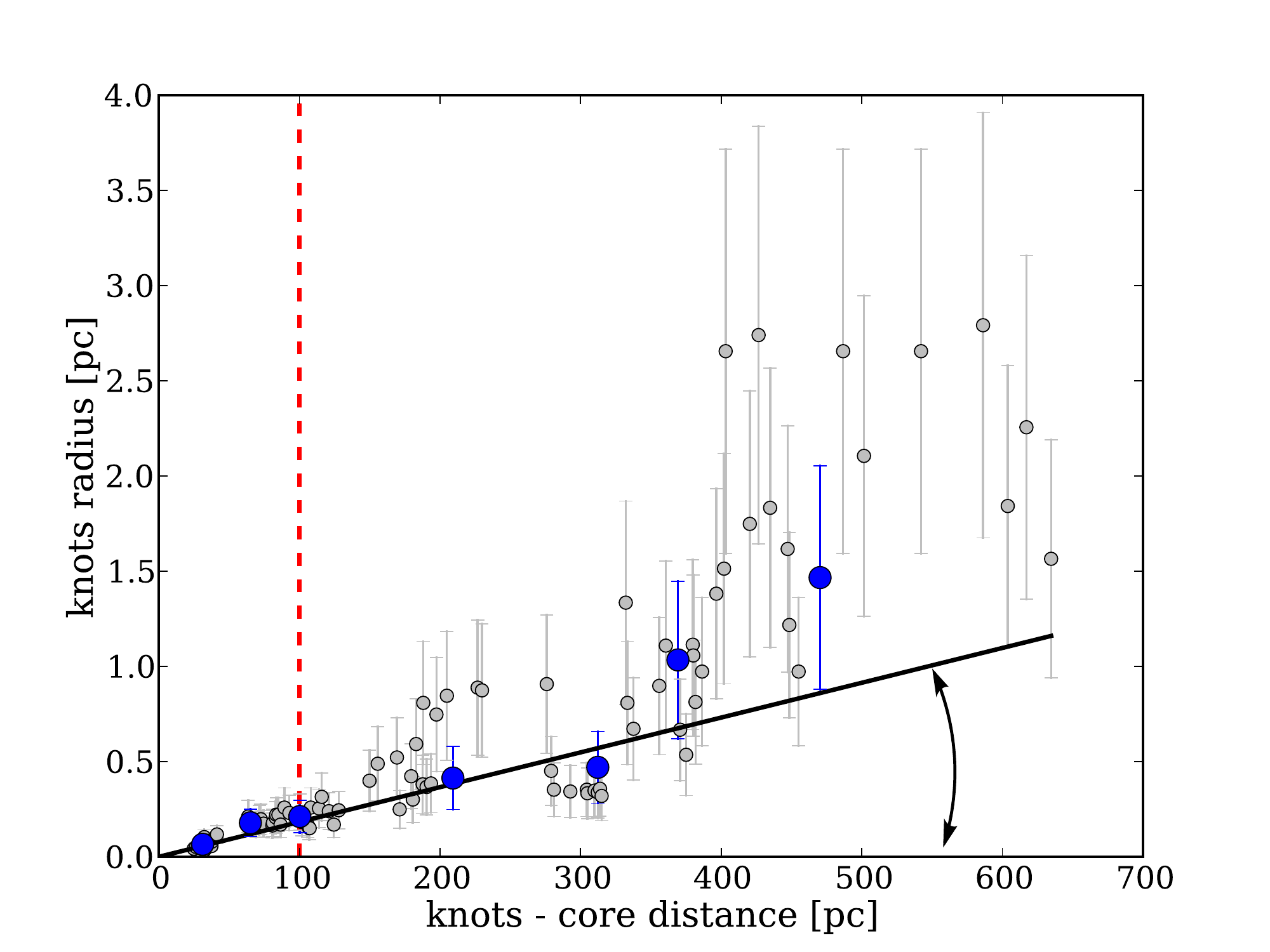}
\put(-53,40){\color[rgb]{0,0,0}\makebox(0,0)[lt]{\begin{minipage}{11cm}\raggedright $\phi$\end{minipage}}}
\caption{Core distances and radius of radio knots observed by MOJAVE for an angle with the line of sight $\theta$ of 1.38$^{\circ}$. {Grey dots} show the referenced knots from August 18, 1997 to March 5, 2011.  {Blue dots} show radio knots of December 26, 2009. The {black line} is a linear regression that we use to characterize the knots expansion angle $\phi$. The {red dashed line} at 100 pc marks the length of the simulated jet.}
\label{fig:ouv_jet}
\end{center}
\end{figure}

To determine the distance to the core and the size of radio knots, we considered all VLBI knots observed by MOJAVE from August 18, 1997 to March 5, 2011 referenced in the Table 4 of \cite{MOJAVE}.
The knot-core distance $ D_{k}$ was deduced from the apparent (projected) knot-core distance $D_{k,proj}$ as $D_{k} = D_{k,proj}/ \sin \theta$, and assuming spherical knots and small observation angle such that  $R_{k} = R_{k,proj}$. The evolution of the radius with the core distance can be fitted by a simple law of proportionality $R_{k} \propto D_{k}$ with a high level of confidence ($\chi^2/dof = 0.98$). We can then estimate the half-opening angle of the knot expansion $\phi$ from a linear regression in the knot radius as a function of their distance: 
\begin{equation}
\phi = \arctan \left(\frac{R_k}{D_k}\right) = 0.10^\circ
.\end{equation}

\begin{figure}[h]
\begin{center}
\includegraphics[width=7.5cm]{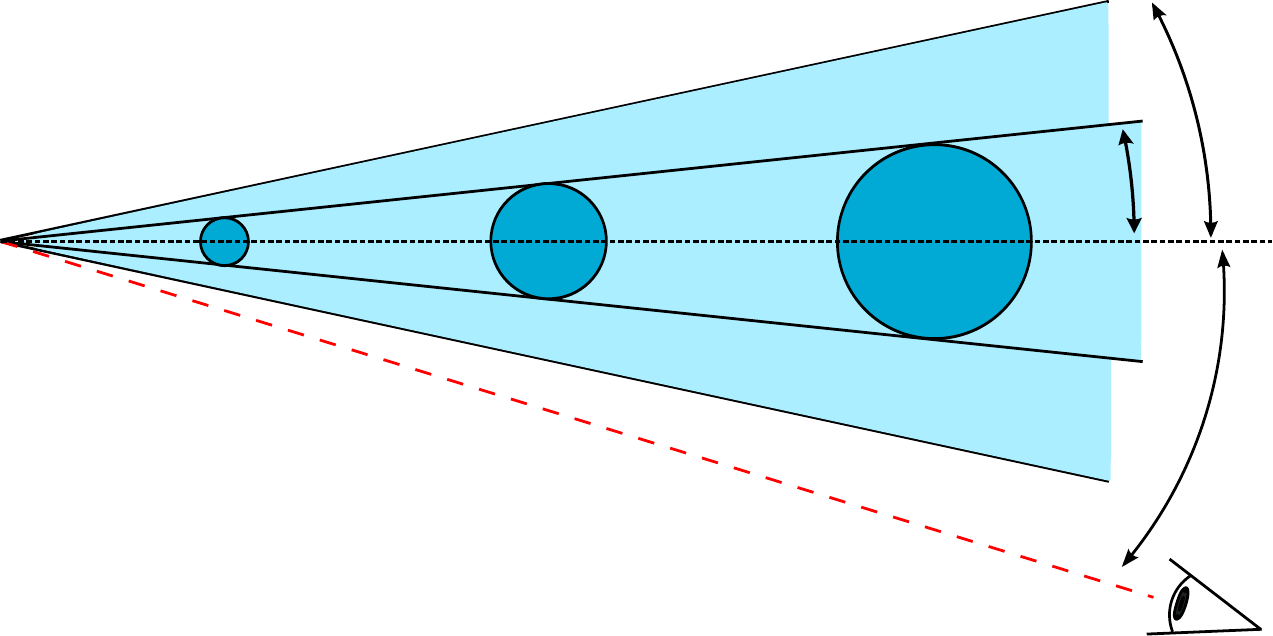}
\put(-20,72){\color[rgb]{0,0,0}\makebox(0,0)[lb]{\smash{$\phi$}}}
\put(-10,85){\color[rgb]{0,0,0}\makebox(0,0)[lb]{\smash{$\alpha$}}}
\put(-10,30){\color[rgb]{0,0,0}\makebox(0,0)[lb]{\smash{$\theta$}}}
\caption{Sketch showing the knot expansion angle $\phi$, the opening jet angle $\alpha,$ and the angle with the line of sight $\theta$. Not to scale.}
\label{jet2}
\end{center}
\end{figure}

This yields $\phi < \alpha < \theta$, and the jet does not strictly point in the observer direction. Because the observation angle $\theta$  is very close to the jet aperture angle $\alpha$ ($\theta - \alpha < 1^\circ$), we assume that radiation transfer formulae are still valid for such a small misalignment.

This geometry describes two jets with two different opening angles, one innermost jet corresponding to the knot expansion and an external jet with a wider opening angle (see Fig. \ref{jet2}). This interpretation is consistent with the observations of \cite{Perucho} and \cite{MOJAVE}, who deduced a thin ribbon-like structure embedded within a broader conical outflow in several VLBI quasars.

\section{SED modelling and discussions}
\label{section:modelling}

\begin{figure*}[t]
\begin{center}
\includegraphics[width=18cm]{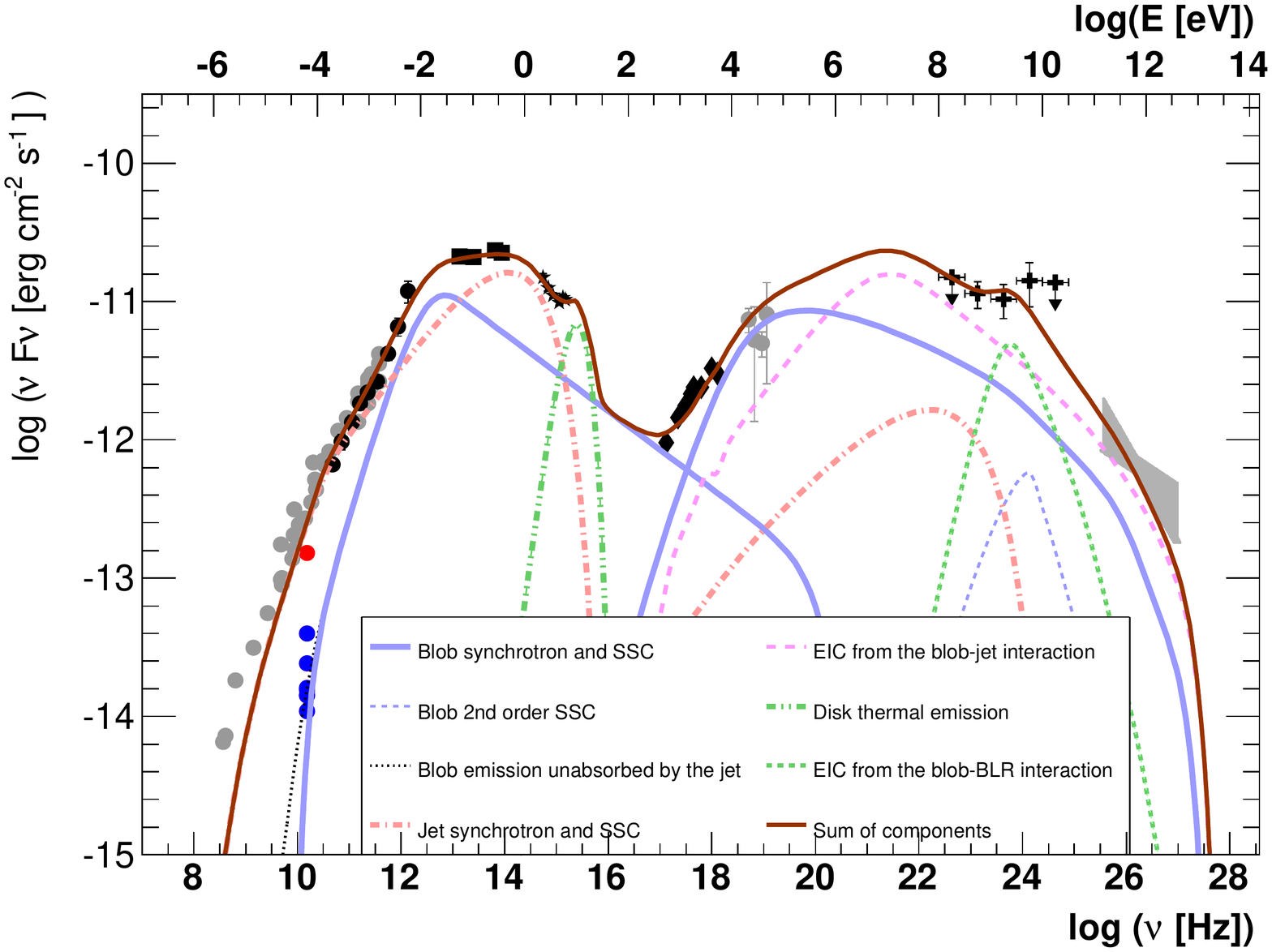}
\caption{Multi-component SSC modelling. Back data: simultaneous or quasi-simultaneous observations, grey data: non-simultaneous observations, blue dots: VLBI radio knots, red dot: VLBI radio core.
Values of physical parameters are given in Table \ref{params}. The EBL absorption based on the model of \cite{EBL} is taken into account in this modelling. Because of the low redshift, the EBL unabsorbed and absorbed spectra are very close, we do not show the unabsorbed spectra for sake of visibility on the SED.}
\label{fig:SED_total}
\end{center}
\end{figure*}

From the modelling of the different components presented in Sect. \ref{multicomponent} and the constraints on physical parameters deduced in Sect. \ref{radio}, we can now generate the SED of Ap Lib in Fig. \ref{fig:SED_total} with the parameters of Table \ref{params}. As we can see in Fig. \ref{fig:SED_total}, the unusual position of the jet and blob synchrotron peaks provide a very good fit of the lower part of the SED, in addition to a good reproduction of the HE and VHE spectra.

\begin{table}[h]
\caption{\label{params} Physical parameters used for the SED modelling presented in Fig. \ref{fig:SED_total}. $L$ is the size of the jet, $D_{blob-BH}$ is the distance between the blob and the central engine.}
\centering
\begin{tabular}{ccl}
        \hline \hline
        \noalign{\smallskip}
    Blob parameters & Value & Unit\\ \hline
    \noalign{\smallskip}
    $\delta_b$ & $22$ & $-$\\ 
    $\theta_b$ & $1.4$ & deg\\ 
    $K_b$ & $2.0\times 10^{5}$ & cm$^{-3}$\\ 
    $n_1$ & $2.0$ & $-$\\ 
    $n_2$ & $3.6$ & $-$\\ 
    $\gamma_{min/b}$ & $600$ & $-$\\
    $\gamma_{max/b}$ & $4.0\times 10^{6}$ & $-$\\
    $\gamma_{break/b}$ & $8.0\times 10^{2}$ & $-$\\
    $B_b$ & $6.5\times 10^{-2}$ & G\\
    $R_b$ & $6.2\times 10^{15}$ & cm\\ \hline
        \noalign{\smallskip}
    Jet parameters &  & \\ \hline
    \noalign{\smallskip}
    $\delta_{jet}$ & $10.0$ & $-$\\ 
    $K_{1,jet}$ & $50.0$ & cm$^{-3}$\\ 
    $n_{jet}$ & $2.0$ & $-$\\  
    $\gamma_{max/jet}$ & $1.65\times 10^{4}$ & $-$\\
    $B_{1,jet}$ & $8.0\times 10^{-2}$ & G\\
    $R_{1,jet}$ & $2.5\times 10^{16}$ & cm\\
    $L$ & $100$ & pc\\
    $\alpha$ & $0.4$ & deg\\
    $D_{blob-BH}$ & $7.9\times 10^{18}$ & cm\\
    $nb_{slices}$ & $50$ & $-$\\ \hline
    \noalign{\smallskip}
    Nucleus parameters &  & \\ \hline
    \noalign{\smallskip}
    $T_{disk}$ & $3.2\times 10^{4}$ & K\\
    $L_{disk}$ & $5.0\times 10^{43}$ & erg.s$^{-1}$\\
    $R_{BLR}$ & $7.9\times 10^{18}$ & cm\\
    $\tau_{BLR}$ & $3.5\times 10^{-2}$ & $-$\\ \hline
    
 \end{tabular}
\end{table}

From modelling the UVOT data with a black body, we deduce a temperature of $3.2\times 10^4$ K and a luminosity of $5.0\times 10^{43}$ erg.s$^{-1}$, which match the usual features of an AGN accretion disk.

The flux density of each VLBI radio knots (blue dots in Figs. \ref{fig:ouv_jet} and \ref{fig:SED_total}) is well reproduced by the synchrotron emission of the blob unabsorbed by the jet, and the core flux density (red dot) by that of the pc-jet.
This gives us an overall view of the emission processes. The radio core emission comes from the base of the jet. We cannot see the radio emission of the emerging blob because of the thickness of the core. But at a large distance from the core, the jet becomes optically thin in radio, and moving blobs can be seen as radio knots. Thus, the SED modelling supports the idea that the simulated VHE blob will become a radio knot. According to this model, the jet becomes radio transparent at 15.4 GHz at about 6.5 pc from the core. This distance projected on the sky plane corresponds to a gap fro, the core of
 $0.24$ mas. The average separation of the closest radio knots from the core reported by \cite{MOJAVE} is roughly 0.85 mas, which is consistent with our proposal. In the following part of this section we examine different aspects for characterizing the link between the VHE blob and the radio knots.

\subsection{Cooling time}

The characteristic synchrotron cooling time $\tau_{cool}(\gamma)$ is
\begin{equation}
\label{eq:cooling}
\tau_{cool}(\gamma) = \frac{3 m_e c}{4 \sigma_T U'_B~\beta^2 \gamma}
,\end{equation}
with $\sigma_T$ the Thomson cross-section and $U'_B = B^2/8\pi$ the magnetic energy density.\\

We obtain a characteristic synchrotron cooling time of 12.8 hours for the very high energy of the blob ($\gamma_{max} = 4.0\times10^6$). This time is much shorter than the year measured between the various radio knot ejections.
In a purely ballistic scenario, where a high-energy blob is emitted  with an injection time of $t_{inj} = R_b/c \simeq 57$h, then moves along the jet and emerges as a radio knot, we should detect very high energy flares only during the rare ejections of knots. However, the source appears stable at very high energy over years, which indicates that the particle acceleration must be continuous.

The 15.4 GHz detection frequency of the radio knots is very close to the minimal blob frequency (see Fig. \ref{fig:SED_total}), so we assume that the mean Lorentz factor $\gamma_r$ of the particles emitting in radio corresponds to the blob $\gamma_{min}$ (see Table \ref{params}). The associated synchrotron cooling time is $\tau_{cool}(\gamma_r) = 9.7$ years. \\
However, the knots with apparent movement detected by MOJAVE occur between 209 pc and 470 pc to the core on December 26, 2009 (see Fig. \ref{fig:ouv_jet}), which means assuming a constant velocity that they were ejected about 700 to 1500 years ago, a time much longer than the cooling time. Thus the particle acceleration should occur not only at the base of the jet, but also constantly at large distances from the core.

The poor angular resolution at very high energy does not allow determining where the gamma radiation zone lies. Two different scenarios can be considered:

{(1)} Either the gamma emission zone is stationary at the base of the inner jet and radio knots are created during large-scale perturbations due, for example, to a shearing between inner and extended jet.

{(2)} Or the gamma emission zone propagates through the jet at the radio knot velocity and continues to radiate at high energy above the pc scale.

The radio cooling time estimate, which requires an efficient acceleration mechanism at large distances, tends to favour the second scenario.
Information on the variability pattern at high energy on long time scales is needed to disctinguish between the two options.
The detection of a gamma flux periodicity corresponding to the radio knot periodicity ($\sim$ year) would support the second one, but this is beyond the scope of the present study.

In the following part we study the jet energies in detail to characterize this continuity between the radio knots and the  VHE blob.

\subsection{Cold particle population}
\label{cold}

Figure \ref{fig:ouv_jet} suggests a linear expansion of radio knots with their propagation along the jet. Assuming that the high-energy emerging blob follows the same expansion as the observed radio knots, we can approximate its expansion velocity with the Alfven velocity $v_{A,b}$  as
\begin{equation}
v_{A,b} = \beta_{b} \tan\phi \simeq 1.74 \times 10^{-3} \text{c}
.\end{equation}

The Alfven velocity depends on the mass density $\rho$ and the magnetic field $B$, we can deduce the blob mass density $\rho_b$ from Table \ref{params}:
\begin{equation}
\rho_b = \frac{1}{4\pi}\left(\frac{B_b}{v_{A,b}}\right)^2 \simeq 1.2\times 10^{-19} \text{g.cm}^{-3}
.\end{equation}

That the proton mass is about a thousand times heavier than the electron mass implies that the blob mass density is widely dominated by protons. We can deduce the proton density $N_{b,p}$, assuming that the environment is electrically neutral, and neglecting any significant cold population composed of e$^-$/e$^+$ pairs, which would annihilate very quickly,
\begin{equation}
N_{b,p} \simeq \frac{\rho_b}{m_{p}} \simeq 7.4\times10^4 \text{cm}^{-3}
.\end{equation}

We determine the density of non-thermal electrons $N_{b,e,NT}$ by integrating over the non-thermal energy distribution spectrum $N_e(\gamma)$:
\begin{equation}
N_{b,e,NT} = \int_{\gamma_{min}}^{\gamma_{max}}{ N_{b,e}(\gamma) d\gamma} \simeq 1.8\times 10^2 \text{cm}^{-3}
,\end{equation}

with values of $\gamma_{min}$ and $\gamma_{max}$ presented in Table \ref{params}. The cold electron population for this neutral environment is
\begin{equation}
N_{b,e,cold} = N_{b,p} - N_{b,NT} = 7.4\times10^4 \text{cm}^{-3}
.\end{equation}
 
In this way, we characterize the three blob components, namely the non-thermal electrons, the cold electrons, and the cold protons. The cold population of electrons appears to represent $99.8\%$ of the total electron population.
\\

The same calculation for the pc-jet at its base provides an Alfen velocity $v_{A,j} = 6.9 \times   10^{-3}$c corresponding to the mass density $\rho_j = 7.7\times 10^{-21}$g.cm$^{-3}$. The proton density is $N_{j,p} = 4.6\times10^3$cm$^{-3}$, very close to the cold electron density $N_{j,e,cold} = 4.5\times10^3$cm$^{-3}$ , which represents $97.2\%$ of the total electron population $N_{j,e,tot}$. The calculated non-thermal electron density at $N_{j,e,NT} = 1.3\times10^2$cm$^{-3}$.

In both the blob and the pc-jet, the cold particle density widely
dominates the density of non-thermal particles, which does not contradict our assumption of continuous evolution between the VHE blob and the radio knots.
The wider aperture of the pc-jet results in a cold particle density lower than that of the blob.

\subsection{Energy budget}
\label{section:energy_budget}

Now we can examine energetic questions. Following the description of \cite{Power}, we can determine the power of various components through a section $\pi R^2$ of the jet by the equation
\begin{equation}
P_i = \pi R_i^2 \Gamma_i^2 c(U'_i)
\label{P}
,\end{equation}

where $U'_i$ is the energy density of the i$th$ component in the comoving frame. The different components of power represent the magnetic field ($P_B$), the non-thermal electron population ($P_{e,NT}$), the cold electron population ($P_{e,cold}$), the cold proton population ($P_{p}$), and the radiation ($P_r$) contribution.

The associated energy densities are\\

\noindent $U'_B$ (see Eq. \ref{eq:cooling}),

\begin{equation}
U'_{e,NT} = m_e c^2 \int_{\gamma_{min}}^{\gamma_{max}}{\gamma N_e(\gamma) d\gamma}
,\end{equation}
where $N_e(\gamma)$ is the non-thermal energy spectrum,

\begin{equation}
U'_{e,cold} = m_e c^2 N_{e,cold} \\ \text{   and   }\\ U'_{p} = m_p c^2 N_p
,\end{equation}
with the cold densities $N_{e,cold}$ and $N_p$ calculated in Sect. \ref{cold},

\begin{equation}
U'_{r} = \frac{4\pi}{c}\int_{\nu'_{min}}^{\nu'_{max}} I'_{\nu'} d\nu'
,\end{equation}
where $I'_{\nu'}$ is the surface intensity.

We calculated the powers of the blob and of the pc-jet separately. The pc-jet power considered is the power estimated at the jet base. The blob radiation density is composed of densities radiated by synchrotron, SSC, second-order SSC, EIC from the BLR, and EIC from the pc-jet ($U'_{b,r} = U'_{b,syn}+U'_{b,ssc}+U'_{b,ssc2}+U'_{b,eicBLR}+
U'_{b,eicJ}$). The jet radiation density is the sum of energy densities radiated by synchrotron and SSC ($U'_{j,r} = U'_{j,syn} +U'_{j,ssc}$).

\begin{table}[h]
\caption{\label{table:energy}Powers of the different components of the Ap Lib total jet expressed in $\log($P$~$[erg.s$^{-1}])$}
\centering
\begin{tabular}{lcccl}
        \hline \hline
        \noalign{\smallskip}
    Power & Blob & Jet & Total\\ \hline
    \noalign{\smallskip}
    Radiation & 42.7 & 41.7 & 42.7\\ 
    Magnetic & 40.9 & 41.2 & 41.4 \\
    Cold electrons & 43.5 & 42.6 & 43.5\\
        Non-thermal electrons& 43.9 & 42.0 & 43.9\\
    Protons & 46.8 & 45.5 & 46.8\\ \hline
 \end{tabular}
\end{table}

The source is dominated by the kinetic power of particles, mainly in the cold protons of the blob, which represent $99.8 \%$ of the total power (see Table \ref{table:energy}). The non-thermal electrons slighty dominate the cold electron population in the blob, while it is the opposite in the jet, but the two cases show almost an equipartition between cold and non-thermal electrons. The magnetic field appears far below the equipartition with the non-thermal electrons and the radiation in the blob, while this equipartition is almost achieved in the longer lifetime jet.\\

\begin{figure}[h]
\begin{center}
\includegraphics[width=8cm]{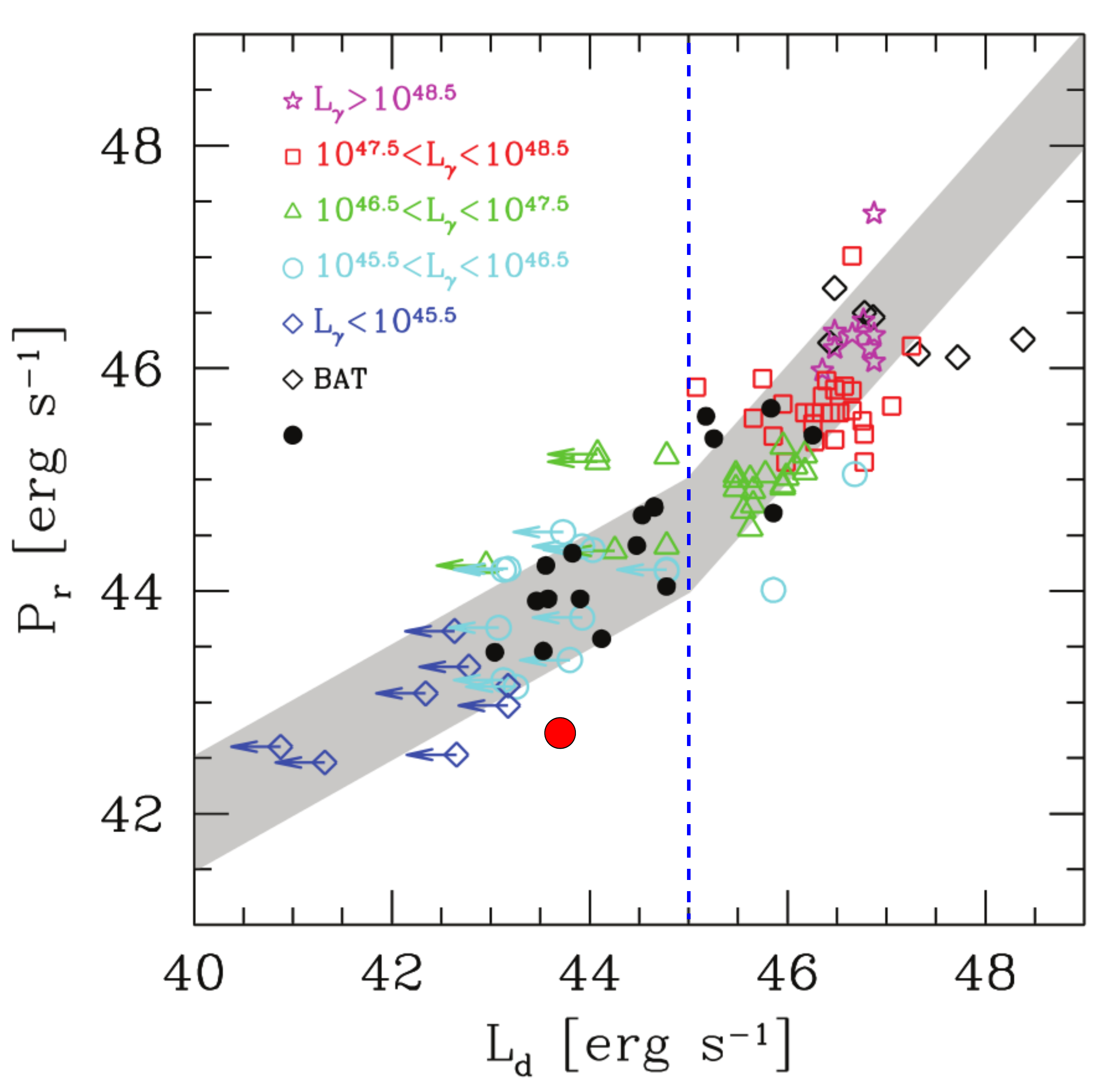}
\put(-80,200){\color[rgb]{0,0,1}\makebox(0,0)[lb]{\smash{{Standard disk}}}}
\put(-170,45){\color[rgb]{0,0,1}\makebox(0,0)[lb]{\smash{{ADAF}}}}
\put(-164,129){\color[rgb]{0,0,0}\makebox(1,0)[lb]{\smash{\small\textbf{Ghisellini et al.}}}}
\put(-164,119){\color[rgb]{0,0,0}\makebox(1,0)[lb]{\smash{\small\textbf{2011}}}}
\put(-126,60){\color[rgb]{1,0,0}\makebox(1,0)[lb]{\smash{\tiny\textbf{Ap Lib.}}}}
\caption{Blazar population in terms of disk luminosity $L_d$ and jet radiation power $P_r$ (figure in \cite{Ghi2011}). The large red point represents the location of Ap Lib. The blue dotted line represents the separation between the two accretion regimes at $L_d \simeq 10^{45}$ erg.s$^{-1}$.
 The radiation power considered here for Ap Lib is the sum of the blob and jet radiation powers.}
\label{fig::ghi2011}
\end{center}
\end{figure}
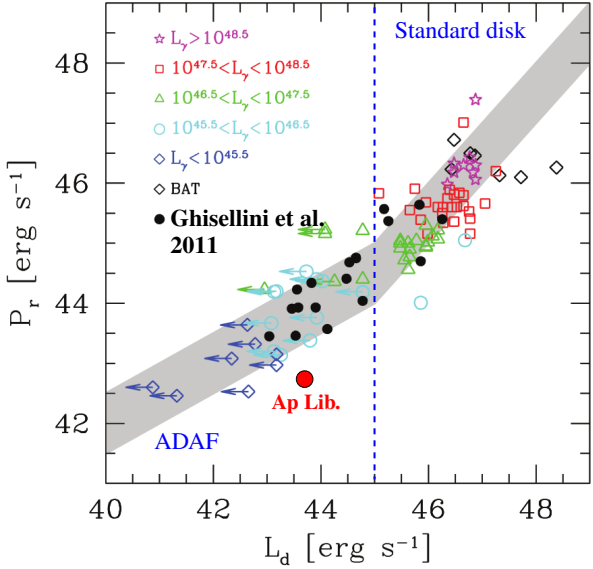

Figure \ref{fig::ghi2011} adapted from \cite{Ghi2011} allows highlighting the peculiarity of Ap Lib. The right part of the graph ($L_d > 10^{45}$erg.s$^{-1}$) shows the very bright sources, mostly of the FSRQ type. In this part, $P_r$ is proportional to the disk luminosity $L_d$. Assuming that $P_r$ is also proportional to the jet power $P_{total}$, this means that for these sources the accretion efficiency, hence the accretion regime, remains constant. This is not the case anymore for disk luminosities below $10^{45}$ erg.s$^{-1}$ where the radiation power instead
evolves in $\sqrt{L_d}$. 
These two slopes of the accretion regime correspond to two different accretion disks models. On the one hand, the most luminous sources are described by standard disks, on the other hand, fainter sources are more consistent with disk models called advection-dominated accretion flow (ADAF), or radiative inefficient accretion flow (RIAF).

The jet radiation power should normally place Ap Lib as BL Lac in the ADAF regime. However, although it is below this critical disk luminosity, Ap Lib seems to better follow the FSRQ type accretion regime, which is located in the prolongation of the grey FSRQ band, as could be an extremely faint FSRQ.
One possibility could be that Ap Lib is an old FSRQ with an accretion disk that became weaker and drifted toward the ADAF region (of low luminosities), but stayed meta-stable in the standard disk regime. 
This idea is supported by all the signs highlighted in our present study that are typical for an FRII type source such as the superluminal radio knots, the low aperture of the jet, and the large kinetic power dominance. In the following we also check whether the accretion efficiency is consistent with an FRII source.

\subsection{Accretion efficiency}

Since the kinetic protons power widely dominates all other components, the mass outflow rate $\dot M_{out}$ can be expressed as
\begin{equation}
\dot M_{out} = \frac{1}{c^2}\left(\frac{P_{j,p}}{\Gamma_j} + \frac{P_{b,p}}{\Gamma_b} \right) \simeq 6.0\times10^{24} g.s^{-1}
.\end{equation}

Taking in account that $\dot M_{out} \leq  \dot M_{in}$, with $\dot M_{in}$ the mass inflow rate (or mass accretion rate), allows placing an upper limit on the accretion efficiency $\eta = L_d /(\dot M_{in} c^2)$,
\begin{equation}
\eta \leq \frac{L_d}{\dot M_{out} c^2} = 9.30 \times10^{-3}
.\end{equation}

This efficiency is lower than the one generally used for quasars of $\eta \simeq 0.1$ \citep{accretion_efficiency}. For $\dot M_{out} \simeq \dot M_{in}$, the accretion efficiency is already at the limit between ADAF and standard disk regime of $\eta \simeq 0.01$ proposed by \cite{ADAF}.\\

\cite{contraintes_phys_blazar} classified different blazar types in terms of $\eta_{crit} = L_{d}/L_{Edd}$, which can be identified as $\dot M_{in} = \dot M_{crit}$, with $\dot M_{crit} = L_{Edd}/c^2$ the critical mass inflow rate.
The Eddington luminosity of Ap Lib is\begin{equation}
L_{Edd} = \frac{4\pi G M_\bullet m_pc}{\sigma_T} \simeq 1.3\times10^{38}\frac{M_\bullet}{M_\odot} \text{ erg.s}^{-1}
,\end{equation}

and its black hole mass is estimated as $M = 10^{8.4\pm 0.06} M_\odot$ \citep{BH_mass}, which gives $L_{Edd} = 1.60\pm 0.07\times 10^{46} erg.s^{-1}$. 
This corresponds to $\eta_{crit} \simeq3.1\times10^{-3}$, very close to the limit value of $3\times10^{-3}$ between BL Lac and FSRQ proposed by \cite{contraintes_phys_blazar}. \\

As shown in Sect. \ref{section:energy_budget}, the value $\eta_{crit}$ does not allow distinguishing the various luminous FSRQs because the accretion efficiency beyond $L_d \simeq 10^{45}$ erg.s$^{-1}$ seems constant. However, it makes a difference between BL Lacs and FSRQs, and also between various BL Lac sources.

These estimates of the accretion efficiency suggests that the source is in a peculiar accretion mode intermediate between ADAF and standard disk. Moreover, the radio features analysed in previous sections as well as the extended X-ray emission studied by \cite{X_jet} show an intermediate blazar between FR I and FR II types. In all aspects, Ap Lib corresponds to an intermediate case in the global blazar dichotomy scheme, which usually ties FRSQ with FR II and standard disk on the one hand and BL Lac with FR I and ADAF disk on the other hand.

\section{A specific type of LSP blazars?}
\label{section:classification}

Simple one-zone SSC scenarios are often able to reproduce the SED and spectra of HBL sources reasonably well. In this type of blazar, the pc-jet, in which is embedded the VHE emitting blob, mainly contributes to the radiative output at low radio frequencies. At high energies, the contribution of this jet is weak or even negligible, for instance in the well-known HBL PKS 2155-304 \citep{PKS2155_MWL, PKS2155}. Conversely, the case of Ap Lib appears very uncommon in the sense that the pc-jet seems to dominate the SED in the radio and optical ranges and significantly contributes to the VHE part of the spectrum. Moreover, the strong radiation of the accretion disk places the source between the two populations of  BL Lac and FSRQ. These properties make Ap Lib a rather unique object for the moment.

 To investigate whether Ap Lib is a special type of blazar or if it could represent a new still unrecognized family of sources, we have identified three main characteristics of its SED as follows: 

{1)} A broken radio-mm spectrum around 250 GHz.

{2)} An X-ray positive slope.

3)\hspace{1mm}A high-energy bump very much wider than the synchrotron bump with a relatively flat Fermi spectrum at $\nu \leq 10^{15}$ GHz and a peak luminosity close to the peak synchrotron luminosity. \\

We interpret the first characteristic as a superposition of two synchrotron bumps, one from a strong jet and one from a compact energetic blob embedded in it. The second characteristic is the signature of the inverse-Compton component dominance from soft X-rays to high energies. The third characteristic is explained by a strong external inverse-Compton emission on the jet and disk radiation.

Characteristics 2 and 3 are relatively common to describe LBL spectra. 
However, characteristic 1 is very peculiar. Following the description presented in Fig. (2) of \cite{blazar_enveloppe}, this change of slope at 250 GHz would be a key point for recognizing a source as a misaligned radio galaxy. But considering that Ap Lib is a blazar with a very low angle with the line of sight (see Sect. \ref{radio_core}), we explain this feature by an unusual strong pc-jet emission and not by a strong misalignment. Therefore this radio-mm change of slope can be seen as a major point for characterizing the source.

To see how common these three features are, we considered the SED of 105 blazars obtained simultaneously by Planck, Swift, and Fermi reported by \cite{giommi}. Indeed, twenty of these 105 sources fulfil the three characteristics. All of them are classified as LSP BL Lac or FSRQ. 
Almost all sources with a spectral radio-mm slope change around 250 GHz have a positive X-ray slope and a wide Compton bump. 
Accordingly, we name here the sources that fulfil these three criteria BSRQs for broken-spectrum radio quasars, in which the emission of the radio VLBI base is relatively high and the blob synchrotron emission reaches maximum at very low-frequency. These two components significantly contribute and interact to produce the non-thermal emission from radio up to the VHE range. 

Thus Ap Lib apparently is not the only member of its type and can be seen as a prototype of VHE blazars, which could be seen as a radiatively extremely faint FSRQ, intermediate between pure FSRQs and pure BL Lacs. In these types of specific VHE blazars, the whole SED is no longer dominated by a one-zone VHE blob as in the case of HBL sources, and additional contributions from the disk and the jet become significant even at very high energies. However, a deeper analysis, which is beyond the scope of the present paper, is necessary to better characterize and identify these specific types of LSP blazars and describe the transition between them and other blazar sources.

\section{Conclusion}

We confirmed that a simple leptonic model dominated by the SSC emission of a single VHE compact zone (or blob)  is unable to reproduce the SED of the blazar Ap Lib. We showed that additional contributions related to the pc-jet and disk radiation appear to play a significant role, even at very high energies. This is consistent with the presence of an unusual strong jet emission, reinforced by the observation at larger scale of an extended X-ray jet. To describe these different components self-consistently, we developed a new code that includes the inverse-Compton emission of the blob particles on the jet radiation and on the disk radiation scattered by the BLR clouds, together with the absorption effects of the blob radiation by the jet. These effects naturally appear in blob-in-jet scenarios, but were previously found to be negligible for the modelling of HBL sources at very high energies. However, specific features of the SED of the LBL object Ap Lib can naturally be interpreted as their signatures and allow constraining their physical parameters. On one hand, the two populations of relativistic particles, in the pc-jet and in the blob, provide a direct explanation for the spectral slope change detected in the radio-mm range, and enable reaching the observed VHE fluxes thanks to the new inverse-Compton component on the jet radiation. On the other hand, the disk radiation and its subsequent scattering by the BLR clouds can simultaneously reproduce the spectral index seen in the UV band with UVOT and the apparent flatness of the Fermi spectrum. 

After identifying the pc-jet as an important component of the global radiative output of the Ap Lib nucleus, we constrained several of its characteristics from available VLBI data.
The observed radio core properties constrain the aperture angle of the pc-jet and the angle with the line of sight, moreover, the velocity and the size of radio knots observed in the jet constrain the knots expansion and the Doppler factor. We deduced a structure with an inner jet embedded in the pc-jet at small viewing angle to the observer. The simulated blob features are consistent with the radio knot observations, but the VHE blob is probably much closer to the core than the radio knots. 
The VHE blob behaves like an emerging  knot in expansion, which can later evolve into some VLBI knot and finally contribute to form the radio jet. The source appears dominated by the kinetic power of particles, mainly by the cold protons of the blob.

The classification of Ap Lib as a BL Lac seems not obvious anymore. Its radio morphology, SED shape, accretion disk luminosity, powers, and X-ray properties correspond to an intermediate case between FSRQs and BL Lacs.
Although Ap Lib is a uncommon object among VHE blazars, the peculiarities found in its SED appear not unique. The spectral index change in the radio-mm range around 250 GHz is found in several other LSP BL Lacs or FSRQs that share other SED similarities with Ap Lib, namely an X-ray positive slope and a large Compton bump. 
 We proposed to name this peculiar case BSRQ for broken-spectrum radio quasar, in reference to this unusual radio-mm index change in the SED and to distinguish them from the standard LBL category, which is composed of FR I type sources without  such a strong pc-jet. An in-depth study of these peculiar BSRQs blazars is necessary to achieve a better understanding of their mechanisms and the classification,
but this is beyond the scope of this paper.

\begin{acknowledgements}
We thank Andreas Zech and H.E.S.S. collaboration members for helpful discussions. This research has made use of data from the MOJAVE database that is maintained by the MOJAVE team \citep{Mojave_thanks}. We acknowledge the use of data and software facilities from the ASI Science Data Center (ASDC), managed by the Italian Space Agency (ASI). Part of this work is based on archival data and on bibliographic information obtained from the NASA/IPAC Extragalactic Database (NED) and from the Astrophysics Data System (ADS) of the SIMBAD database, operated at CDS, Strasbourg, France.
\end{acknowledgements}

\begin{noteadd}
While this article was with the referee, the \cite{Sanchez_2015} redefined the VHE spectrum of Ap Lib with more observation time. They reported a flux of $(8.8 \pm 1.5_{stat} \pm 1.8_{sys}) \time 10^{-12}$ cm$^{-2}$ s$^{-1}$ above 130 GeV , and a spectral index of $\Gamma = 2.65\pm0.19_{stat}\pm 0.20_{sys}$ between May 10, 2010 and May 08, 2011.
Although it includes observations from 2011, this spectrum remains close to the one used here. It does not affect the capacity of the model developed here to reach the VHE range and therefore remains fully compatible with all results discussed above.
\end{noteadd}

\bibliographystyle{aa}
\bibliography{Ap_lib}

\end{document}